\documentclass[twocolumn,aps,prl,reprint,noshowpacs,letterpaper,floatfix]{revtex4-1}
\usepackage{amsmath,amssymb}
\usepackage{txfonts}
\usepackage{bm}
\usepackage{textcomp}
\usepackage[pdftex]{graphicx}
\usepackage[bookmarks=false]{hyperref}

\begin{document}

\title{Enhancement of superconductivity by pressure-induced critical
ferromagnetic fluctuations in UCoGe}

\author{Masahiro Manago}
  \email{manago@scphys.kyoto-u.ac.jp}
\author{Shunsaku Kitagawa}
\author{Kenji Ishida}
  \affiliation{Department of Physics, Graduate School of Science,
  Kyoto University, Kyoto 606-8502, Japan}

\author{Kazuhiko Deguchi}
\author{Noriaki K. Sato}
  \affiliation{Department of Physics, Graduate School of Science,
  Nagoya University, Nagoya 464-8602, Japan}

\author{Tomoo Yamamura}
  \altaffiliation[Present address: ]{Institute for Integrated Radiation
  and Nuclear Science, Kyoto University, Kumatori, Japan.}
  \affiliation{Institute for Materials Research,
  Tohoku University, Sendai 980-8577, Japan}

\begin{abstract}
A $^{59}$Co nuclear quadrupole resonance (NQR) was performed on a
single-crystalline ferromagnetic (FM) superconductor UCoGe under pressure.
The FM phase vanished at a critical pressure $P_c$, and
the NQR spectrum just below $P_c$ showed phase separation of the FM and
paramagnetic (PM) phases below Curie temperature $T_\textrm{Curie}$,
suggesting first-order FM quantum phase transition (QPT).
We found that the internal field was absent above $P_c$,
but the superconductivity is almost unchanged.
This result suggests the existence of the nonunitary to unitary
transition of the superconductivity around $P_c$.
Nuclear spin-lattice relaxation rate $1/T_1$ showed the FM
critical fluctuations around $P_c$, which persist above $P_c$ and are
clearly related to superconductivity in the PM phase.
This FM QPT is understood to be a weak first order with critical fluctuations.
$1/T_1$ sharply decreased in the superconducting (SC) state above $P_c$ with a
single component, in contrast to the two-component $1/T_1$ in the FM SC state,
indicating that the inhomogeneous SC state is a characteristic feature of the
FM SC state in UCoGe.
\end{abstract}

\maketitle
Ferromagnetism and superconductivity have been considered to compete and
mutually suppress one another \cite{SovPhysJETP.4.153},
and the coexistence has been only reported in some compounds so far,
where two phenomena arise from different atoms or sites
\cite{PhysRevLett.1.449,PhysicaC.254.151,PhysRevLett.86.5767}.
However, such a generally accepted notion was forced to change after the
discovery of superconductivity in a series of uranium (U)-based ferromagnets,
namely, UGe$_2$ \cite{Nature.406.587}, URhGe \cite{Nature.413.6856},
and UCoGe \cite{PhysRevLett.99.067006}.
The superconducting (SC) phase in these compounds is embedded inside the
ferromagnetic (FM) phase, and spin-triplet pairing is highly anticipated.
Our $^{59}$Co nuclear-quadrupole-resonance (NQR) measurements showed
that ferromagnetism and superconductivity in UCoGe coexist microscopically
\cite{JPSJ.79.023707},
and that the U site is an origin of two phenomena \cite{JPSJ.80.064711}.
One of the attractive features of these systems is that the FM quantum phase
transition can be achieved experimentally with pressure or magnetic field,
and thus, they are excellent systems for studying the relationship between the
FM and SC phases.
It was reported that the reentrant superconductivity in URhGe
\cite{Science.309.1343} and the robustness of superconductivity in UCoGe are
related with the field-induced FM criticality \cite{JPSJ.78.113709}.
The nuclear-magnetic-resonance (NMR) measurements revealed that the reentrant
superconductivity in URhGe is associated with the tricritical fluctuations
induced by the field \cite{PhysRevLett.114.216401,PhysRevB.93.201112}.
We have shown from direction-dependent NMR
measurements in UCoGe that the longitudinal critical FM fluctuations, which are
regarded as amplitude modes of magnons, play an essential role for
superconductivity, and suggested that the FM fluctuations induce spin-triplet
superconductivity with the theoretical model calculations
\cite{PhysRevLett.108.066403,JPSJ.83.073708,JPhysConfSer.449.012029}.
This scenario, which differs from the ordinary electron-phonon coupling in
the BCS model, is consistent with the theoretical work by Mineev
\cite{PhysUsp.60.121} and is supported by recent thermodynamical
measurements and analyses \cite{NatComm.8.14480}.

One of the remaining issues in UCoGe is an understanding of the FM criticality
and its relationship with the superconductivity under pressure.
UCoGe possesses a unique pressure-temperature phase diagram, in which
superconductivity persists in the paramagnetic (PM) region beyond the FM
criticality \cite{JPSJ.77.073703,PhysRevLett.103.097003,PhysRevB.94.125110}.
This phase diagram implies that the superconductivity is induced by the
fluctuation related to a quantum critical point (QCP).
In the case of antiferromagnetic (AFM) instability,
SC phases are widely observed around the AFM QCPs
in Ce-based heavy-fermion superconductors and iron-based superconductors.
By contrast, the relationship between the FM QCP and the superconductivity is
not straightforward because the first-order quantum phase transition (QPT)
has been anticipated from the theoretical study
of the itinerant ferromagnetism \cite{RevModPhys.88.025006}.
Actually, UGe$_2$ and URhGe exhibit first-order FM transitions by
pressure or magnetic field \cite{PhysRevLett.87.166401,NatPhys.3.460},
and the FM QCP does not exist at zero field.
The first-order FM transition was reported at $P=0$ in
UCoGe \cite{PhysicaC.470.S561,JPSJ.79.023707}, and it is necessary to examine
how the FM criticality relates to the superconductivity in UCoGe.

Another important issue in UCoGe is the identification of a self-induced
vortex (SIV) state in the coexisting phase.
Careful magnetization and superconducting quantum interference device
(SQUID) measurements showed that the
Meissner state is absent although the Meissner-Ochsenfeld effect was observed
\cite{JPSJ.79.083708,PhysRevLett.109.237001}.
We reported from the $^{59}$Co NQR that the FM SC state is inhomogeneous
because two nuclear relaxation components showing SC and non-SC
behaviors were observed below the SC transition temperature $T_\text{SC}$,
and suggested the realization of the SIV state \cite{JPSJ.79.023707}.
However, we could not rule out the possibility that this inhomogeneity
arises from the disorder- or impurity-induced non-SC part.
It is, therefore, crucial to know whether the non-SC component disappears
when the FM state is suppressed by pressure.

In this Rapid Communication, we report that the FM QPT of UCoGe
is weakly first order, and the details of the phase diagram are
different from the case of a second-order QCP.
We also found that the FM fluctuations are enhanced and $T_\textrm{SC}$
increases around $P_c$, indicative of the positive relationship between
the two phenomena.
The strong FM fluctuations persist above $P_c$, and are likely to be
responsible for the SC state in the PM side.
UCoGe is also a member of the FM superconductors showing a first-order FM
transition; however, the discontinuity of the magnetization at the transition
is so weak that the development of the
FM fluctuations is observed, which is a characteristic feature of the
second-order transition.
The NQR measurements above $P_c$ suggest that the PM SC state is homogeneous,
indicating that the whole part of the sample becomes superconducting.
This result leads to a conclusion that the two-component nuclear relaxation in
the FM SC state is not due to the disorder or impurity but is a characteristic
feature in the FM superconductors.

We used a single-crystalline sample with the FM and SC transition temperatures
$T_\text{Curie} = 2.5$ K and $T_\text{SC} = 0.46$ K, respectively.
Details of sample preparation are described in a previous paper
\cite{PhysRevLett.108.066403}.
This sample has a residual resistivity $\rho_0 = 13$ $\mu \Omega$\,cm
along the $b$ axis at ambient pressure \cite{PhysRevLett.108.066403},
and the mean free path is calculated as $l \simeq 700$ \r{A} if we adopt
a rough estimation used in the previous study \cite{PhysRevLett.100.077002}.
This value is larger than the SC coherence length $\xi \simeq 120$ \r{A}
\cite{PhysRevLett.100.077002}.
Hydrostatic pressure was applied using a piston cylinder-type cell with
Daphne oil 7373 as a pressure medium.
Low-temperature measurements were carried out using a $^3$He-$^4$He
dilution refrigerator down to 0.15 K.
The $^{59}$Co NQR was performed without applying static field.
$1/T_1$ detects the FM fluctuations
along the easy ($c$) axis, since the nuclear quantization axis
at the Co site in NQR is almost parallel to the crystallographic $a$
axis \cite{PhysRevLett.105.206403}, and $1/T_1$ is determined with the magnetic
fluctuations perpendicular to the quantization axis.
The RF magnetic field $H_1$ was applied along the $c$ axis, where 
a large signal intensity was obtained.

\begin{figure}
    \centering
    \includegraphics{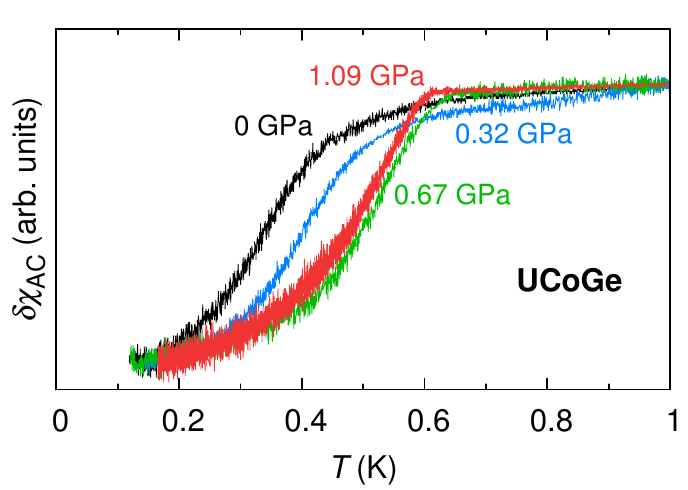}
    \caption{\label{fig:chi}(Color online)
    ac susceptibilities of UCoGe at several pressures measured
    with an NMR coil at zero magnetic field.
    Frequencies are between 5 and 9 MHz.
    }
\end{figure}

Figure \ref{fig:chi} shows the temperature dependence of the ac susceptibility
$\delta \chi_\text{ac}$ of UCoGe at several pressures
measured with an NQR coil.
The $\delta \chi_\text{ac}$ was determined by the change of the
tuning of the $LC$ circuit.
$T_\text{SC}$ slightly increases with increasing
pressure and gets maximum at around 0.67 GPa.
The SC transition width becomes sharper as pressure increases.
These results are qualitatively consistent with the previous studies
with bulk measurements
\cite{JPSJ.77.073703,PhysRevLett.103.097003,PhysRevB.94.125110}.

\begin{figure}
    \centering
    \includegraphics{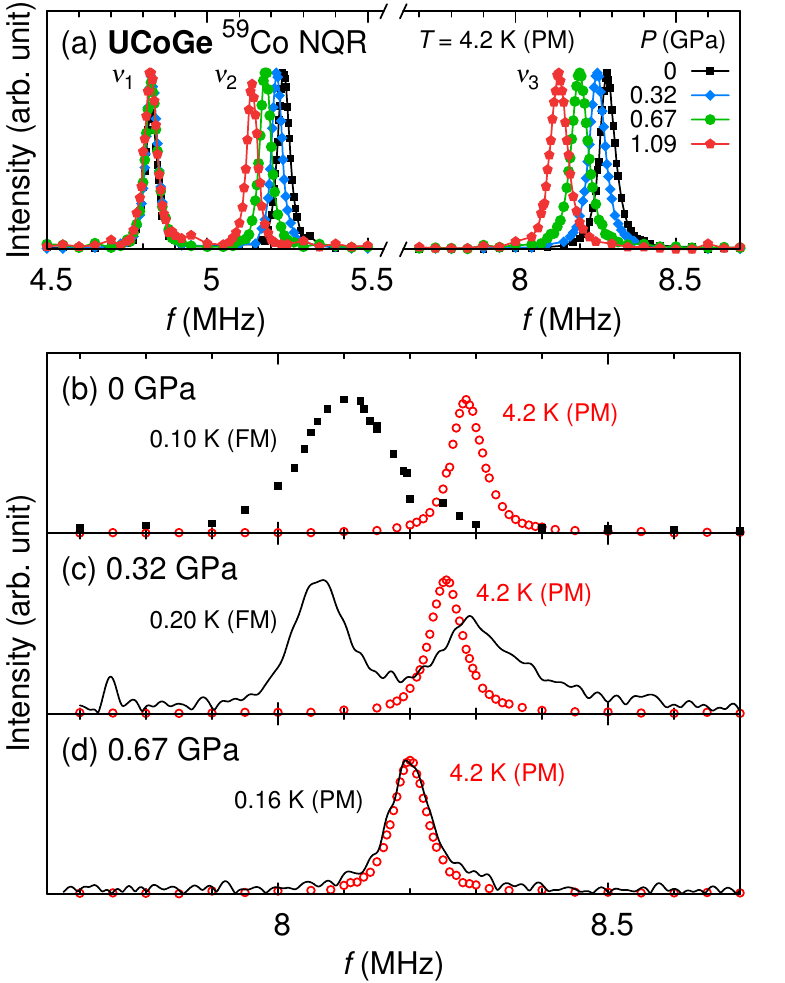}
    \caption{\label{fig:spectrum}(Color online)
    $^{59}$Co NQR spectra of UCoGe under zero field
    (a) at 4.2 K (PM state) arising from $\nu_1$, $\nu_2$, and $\nu_3$
    transitions, and (b--d) at various temperatures at $\nu_3$.
    The spectrum at $P=0$ GPa and $T=0.10$ K shown in (b) was
    from Ref.~\onlinecite{JPSJ.79.023707}.
    }
\end{figure}

Figure \ref{fig:spectrum}(a) shows the pressure dependence of the
$^{59}$Co-NQR spectra at 4.2 K in the PM state.
Three peaks arise from $\pm 1/2 \leftrightarrow \pm 3/2$ ($\nu_1$),
$\pm 3/2 \leftrightarrow \pm 5/2$ ($\nu_2$),
and $\pm 5/2 \leftrightarrow \pm 7/2$ ($\nu_3$) transitions.
The large asymmetric parameter $\eta \simeq 0.52$ makes ratios of
these frequencies far from simple integers.
The spectra slightly shift as pressure increases
because the electric field gradient at the nuclear site changes by the
lattice shrinkage.
The unchanged line width indicates a small inhomogeneity of the applied
pressure.

The temperature variation of the spectra at the $\nu_3$ line at different
pressures is shown in Figs.~\ref{fig:spectrum}(b)--\ref{fig:spectrum}(d).
At ambient pressure, a FM signal appears below $T_\text{Curie}$
because of an internal field at the nuclear site \cite{JPSJ.79.023707}.
The PM signal disappears and only the FM signal was detected at sufficiently
low temperature, suggestive of a homogeneous FM state.
On the other hand, coexistence of the PM and FM signals persists down to the
lowest measurement temperature at 0.32 GPa.
The PM signal becomes broader below $T_\text{Curie}$, which could be ascribed
to the magnetostriction effect \cite{PhysRevB.82.052502}
because the existence of the partial FM regions with slightly different
lattice constants leads to the local stress in the sample.
The two peaks indicate the phase separation of the PM and FM phases
and the first-order FM transition occurs at 0.32 GPa.
Finally, no FM signal was detected at 0.67 GPa.
These results indicate that the FM phase transition of UCoGe is already a
first order at ambient pressure, and is completely suppressed at 0.67 GPa in
our sample, indicating that the FM criticality of UCoGe is classified as
the first-order QPT.
The phase diagram of this sample is qualitatively in good agreement with the
previous studies
\cite{JPSJ.77.073703,PhysRevLett.103.097003,PhysRevB.94.125110},
although $P_c$ is somewhat lower than the values in literature.
This difference may reflect the remarkable sample dependence
of the ferromagnetism of UCoGe \cite{JPSJ.80.084709}.
The unchanged spectrum shown in Fig.~\ref{fig:spectrum}(d) also indicates that
an internal magnetic field was absent in the SC state at 0.16 K within the
experimental resolution.
Thus, a unitary state is realized in the PM SC state.
The time-reversal symmetry has been anticipated
in this phase because the FM transition is absent above $T_\text{SC}$
\cite{PhysRevLett.103.097003}, and the present result verified
the absence of the internal field in the SC state from the microscopic
point of view.
We suggest that the nonunitary-unitary transition
occurs around $\sim 0.5$ GPa in the present sample.

\begin{figure}
    \centering
    \includegraphics{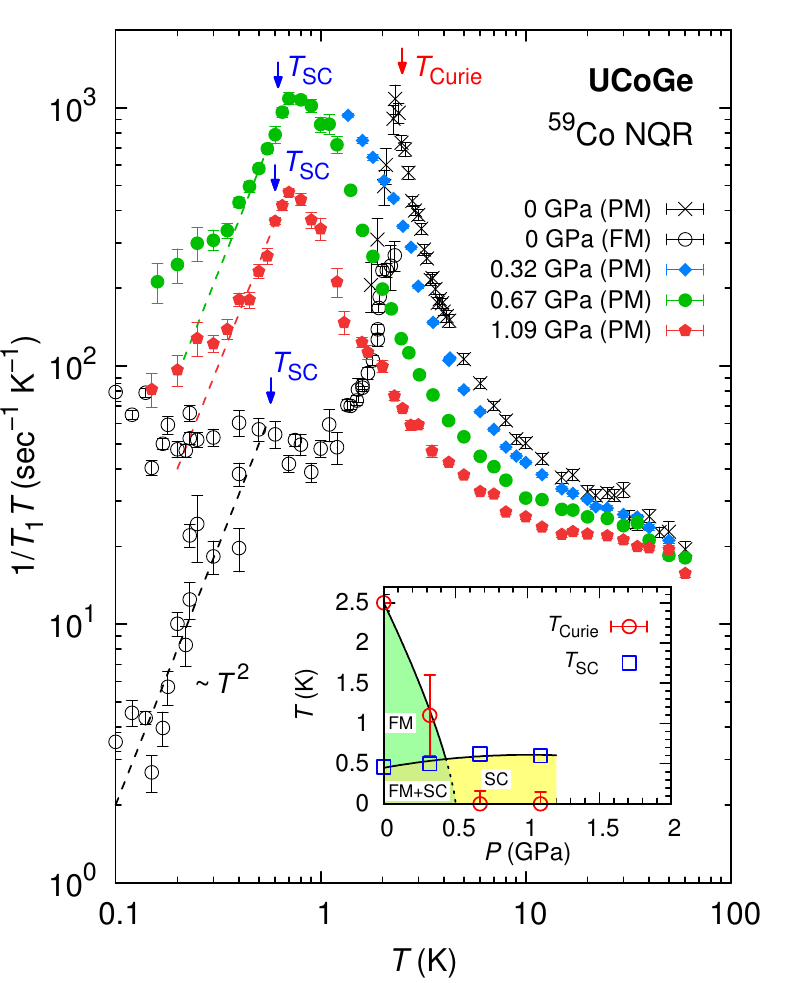}
    \caption{\label{fig:t1}(Color online)
    $^{59}$Co nuclear spin-lattice relaxation rate $1/T_1$ divided by
    temperature $T$ under zero field in UCoGe.
    The result at 0 GPa was from Ref.~\onlinecite{JPSJ.79.023707}.
    The data below 1.4 K lacks for 0.32 GPa owing to the poor NQR intensity.
    $1/T_1$ was measured at the PM site except for 0 GPa below
    $T_\text{Curie}$, and was obtained at the $\nu_3$ line ($\sim 8$ MHz).
    Inset: pressure-temperature phase diagram of UCoGe determined by the ac
    susceptibility and NQR for the present sample.
    }
\end{figure}

Figure \ref{fig:t1} shows the results of nuclear spin-lattice relaxation rate
$1/T_1$ divided by $T$ at several pressures measured at the $\nu_3$ line.
The phase diagram of the present sample determined by the ac susceptibility
and NQR is shown in the inset of Fig.~\ref{fig:t1}.
At ambient pressure, $1/T_1 T$ exhibits strong enhancement around
$T_\text{Curie}$ due to strong FM fluctuations \cite{JPSJ.79.023707}.
When the pressure increases, the peak temperature shifts lower
owing to the suppression of the FM phase, and the FM fluctuations at
$T_\text{SC}$ are strongest at 0.67 GPa.
A clear SC transition was observed even with strong FM fluctuations, and
this is consistent with the scenario that the superconductivity is mediated
by the Ising-type FM fluctuations \cite{PhysRevLett.108.066403}.
At 1.09 GPa, the enhancement of the FM fluctuations becomes weaker and
$T_\text{SC}$ start to decrease, but the enhanced behavior still remains.
The enhancement of $1/T_1$ usually implies a second-order FM QCP, but the clear
first-order transition was observed in the NQR spectrum (Fig.~\ref{fig:spectrum}).
Thus, the FM transition of UCoGe is likely to be weakly first order
near the tricritical point.
We note the possibility that the hyperfine coupling constant might be changed
by applying the pressure.
In such a case, however, $1/T_1 T$ and the Knight shift significantly change
far above the magnetic ordering temperature, which was actually observed in
CeRhIn$_5$ \cite{PhysRevB.92.155147,PhysRevB.65.020504}.
Because $1/T_1T$ is almost invariant at 60 K in UCoGe, the change of
$1/T_1T$ is ascribed to the change of the spin fluctuations by pressure.

The SC state exists in the FM and PM sides
in spite of the first-order FM transition by pressure.
This is owing to the presence of strong FM fluctuations in both sides,
and the discontinuity of the magnetism does not seriously affect the
formation of the Cooper pairs.
This is different from the case of UGe$_2$, where the FM phase vanishes with a
first-order transition and neither critical fluctuations nor
superconductivity was observed above $P_c$ \cite{JPSJ.74.2675}.
In addition, it is shown that $T_\text{SC}$ of UCoGe is the highest at around
the FM criticality.
Thus, the picture that the SC emerges around the FM criticality seems valid in
UCoGe, and this system is similar to the case of the superconductors
observed near the AFM phase.
The remaining issue is whether UCoGe has a
wing structure on pressure-field-temperature phase diagram characteristic to
the quantum itinerant ferromagnet \cite{RevModPhys.88.025006}.

$1/T_1$ gives information about the SC gap structure as well as the magnetic
fluctuations.
Below $T_\text{SC}$, $1/T_1T$ rapidly decreases under all the pressures because
of the opening of the SC gap.
Line-nodal gap behavior was observed in the FM SC state at ambient
pressure in the previous NQR \cite{JPSJ.79.023707}, namely, $1/T_1T \sim T^2$.
Similar line-node behavior is also confirmed with a thermal conductivity
measurement \cite{PhysResInt.2014.454939}.
Recently, it has been proposed that the line node of the SC gap
is protected by the nonsymmorphic space-group symmetry,
and it is expected that this gap structure persists in the PM SC state under
pressure \cite{JPSJ.86.023703}.
However, a deviation from this behavior was observed at 0.67 GPa and 1.09 GPa
in the PM side at low temperatures.
The deviation could be explained by a large residual density of states (DOS) in
the SC state and also supports the nodal SC gap.
The details of the gap structure are masked by this additional relaxation
in the PM SC state.
Since the multigap behavior was reported in the FM SC state at $P=0$
of the recent high-quality samples by the thermal conductivity measurement
\cite{PhysRevB.90.180501}, further measurements are necessary to reveal the
SC gap structure of UCoGe.

\begin{figure}
    \centering
    \includegraphics{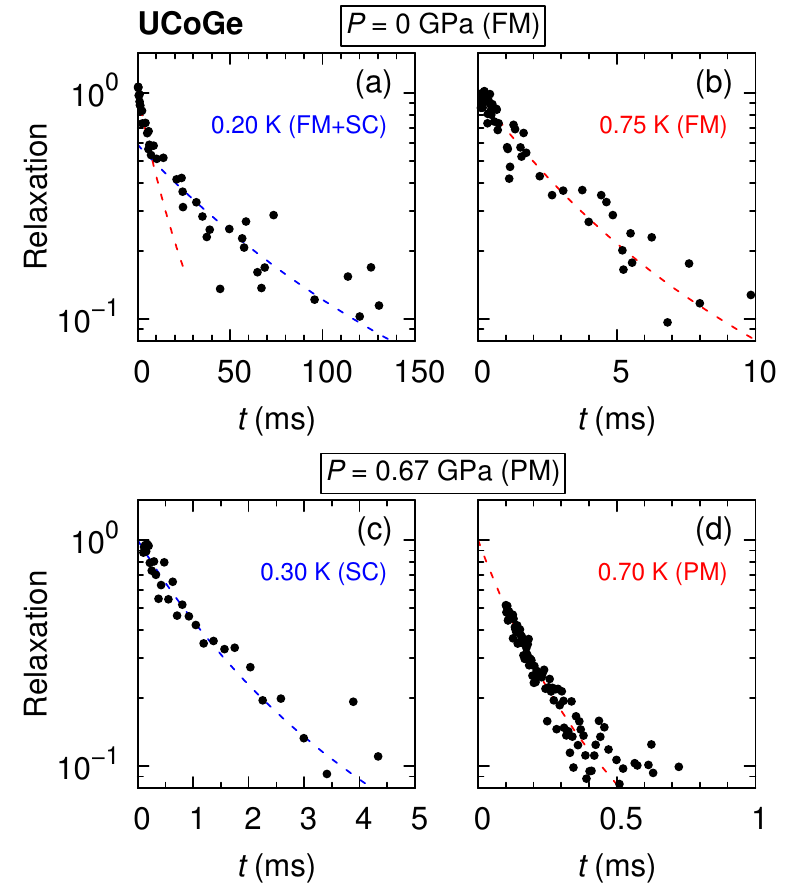}
    \caption{\label{fig:relaxation}(Color online)
    $^{59}$Co NQR relaxation curves of UCoGe at 0 (FM) and 0.67 GPa (PM)
    above and below $T_\text{SC}$.
    These data were measured at the $\nu_3$ line ($\sim 8$ MHz).
    The dashed lines are best fit of the results.
    The curves at ambient pressure denote the previous results
    \cite{JPSJ.79.023707}.
    The single-component relaxation persists in the SC state.
    }
\end{figure}

Figure \ref{fig:relaxation} shows the nuclear relaxation curves at ambient
pressure and 0.67 GPa, measured by $^{59}$Co NQR.
At the FM SC state, the relaxation curve shows two components below
$T_\text{SC}$ as shown in Fig.~\ref{fig:relaxation}(a).
The slower component shows the SC behavior, but the faster component shows the
non-SC behavior (Fig.~\ref{fig:t1}) \cite{JPSJ.79.023707}.
The non-SC part was roughly 50\% in the intensity and did not
show large temperature dependence below $T_\text{SC}$.
On the other hand, a single-component relaxation persists even below
$T_\text{SC}$ in the PM SC state [Fig.~\ref{fig:relaxation}(c)].
These results indicate that the faster component in the FM SC state is an
inevitable feature of UCoGe and originates from the FM phase, and the whole
part of the sample exhibits superconducting in the PM SC state under pressure.
In addition, the temperature-independent fraction of the faster component
would be inconsistent with the nuclear spin diffusion by the presence of the
diffusion center.
Alternatively, we suggest that the faster relaxation could be explained by SIV
owing to the coexistence of the FM and SC phases, as discussed in the previous
paper \cite{JPSJ.79.023707}.
In this scenario, the superconductivity is destroyed at the vortex core,
which results in a normal-metal-like $1/T_1$.
It is theoretically expected that the larger magnetization leads to a larger
residual DOS owing to the partial pair breaking
at the vortex core \cite{JPSJ.82.094711}, and this tendency is consistent with
experiments on the three FM superconductors \cite{CRPhys.14.53}.
Disorder-induced residual DOS behavior was seen in the present $1/T_1$ of UCoGe
in the PM SC state under pressure as mentioned above;
however, such a deviation was absent in the longer component of $1/T_1$
in the FM SC state at ambient pressure.
This difference is also explained by the formation of the
vortex state, because the disorder region in the sample works as a pinning
center of the vortex in the FM SC state and mainly contributes to the
short-$1/T_1$ part.
It should be noted that the large fraction of the non-SC part in the FM SC
state is not evident because the internal field by the spontaneous
magnetization at the FM state is two orders of magnitude smaller than $H_\text{c2}$
along the $c$ axis \cite{JPSJ.79.083708,PhysRevLett.109.237001}, and the
estimation of the SIV region is of the order of the magnetization divided by
$H_\text{c2}$ \cite{JPSJ.82.094711}.
The SIV region would be increased by the presence of a pinning center related
to the disorder.

In addition to the SIV scenario, an interesting scenario to explain the non-SC
fraction is the presence of unpaired electrons related to the spontaneous
charge current resulting from the FM chiral SC state \cite{PhysRevB.97.014519}.
The scanning SQUID measurement revealed that the FM domain wall width
is $\sim 0.1$--1 nm,
and the size of the FM domains, an order of 10 \textmu m, shows no large change
across the SC transition \cite{PhysRevB.90.184501}.
If the FM chiral SC state is realized in UCoGe, it is expected that chiral SC
domains are created and they coincide with the FM domains.
This is the case for the so-called $A$ symmetry expected from the theoretical
works for the interpretation of the NMR results
\cite{PhysRevLett.108.066403,JPSJ.83.073708,JPhysConfSer.449.012029,%
PhysRevB.93.174512}.
The spontaneous charge current would flow at the surface and the domain walls
due to the opposite directions of the multidomain.
With the multidomain structure, a state with finite sum of these currents can
be more stable than a state with cancellation of these currents, and the
Fulde-Ferrell-like supercurrent flows in the opposite direction inside the
domain to cancel the net current in the domain.
This current leads to spatially dependent pair breaking, and thus, this is also
related to the two-component NQR relaxation in our sample
\cite{JPSJ.79.023707} and a large residual specific-heat coefficient
even in a high-quality sample \cite{CRPhys.14.53}.
Although the magnitude of this contribution to these experimental quantities
and the relationship with the SIV state are still unclear, this scenario may
explain the observed non-SC fraction.
Further studies are needed to conclude the origin
of this anomalous behavior in the FM SC state.

In conclusion, $^{59}$Co NQR was performed on the FM superconductor UCoGe
under pressure, and it was revealed that the FM fluctuations are enhanced
around the critical pressure.
This enhancement persists above $P_c$, and is closely related to the emergence
of the SC phase in the PM side.
The phase separation of the FM and PM phases indicates weakly first-order FM
QPT.
The nuclear relaxation curve has a single component in the PM SC state,
which suggests that the fast relaxation in the FM SC state is a
characteristic feature of UCoGe and is closely related to the interplay between
the FM and SC states.

The authors would like to thank Y. Tokunaga, T. Hattori, D. Aoki, Y. Maeno,
S. Yonezawa, A. Daido, Y. Yanase, J.-P. Brison, D. Braithwaite, A. Pourret,
C. Berthier, A. de Visser, J. Flouquet, and V. P. Mineev for valuable
discussions.
One of the authors (MM) is a Research Fellow of
Japan Society for the Promotion of Science (JSPS).
This work was supported by Kyoto University LTM Center, and by
Grant-in-Aid for Scientific Research (Grant No.~JP15H05745),
Grant-in-Aids for Scientific Research on Innovative Areas ``J-Physics''
(Grants No.~JP15H05882, No.~JP15H05884, and No.~JP15K21732), and
Grant-in-Aid for JSPS Research Fellow (Grant No.~JP17J05509) from JSPS.


\end{document}